# XRT Light curves:
# Morphology, Flares and Energy

Guido Chincarini[1,2],
Vanessa Mangano[3], Alberto Moretti[2], Matteo Perri[4], Patrizia Romano[2],
Sergio Campana[2], Stefano Covino[2], GianpieroTagliaferri[2], Lorella Angelini[5],
David Burrows[6], Paolo Giommi[4,7], Julian Osborne[8] and Swift Team

[1] *Università degli Studi Milano – Bicocca, P.za delle Scienze 3, 20126 Milano, Italy*
[2] *INAF – Osservatorio Astronomico di Brera, Via Bianchi 46, 23807 Merate (LC), Italy*
[3] *INAF – IASF Palermo, Via U. La Malfa 153, 90146 Palermo, Italy*
[4] *ASI Science Data Center, Via G. Galilei, 00044 Frascati (RM), Italy*
[5] *Department of Astronomy & Astrophysics, Pennsylvania State University, University Park, PA 16802, USA*
[6] *Department of Physics & Astronomy, University of Leicester, Leicester LE1 7RH, UK*
[7] *Agenzia Spaziale Italiana, Unità Osservazione dell'Universo, Viale Liegi 26, 00198 Roma, Italy*
[8] *NASA/Goddard Space Flicht Center, Greenbelt, MD 20771, USA*

## ABSTRACT

Following a brief introduction we show that the observations obtained so far with the Swift satellite begin to shed light over a variety of problems that were left open following the excellent performance and related discoveries of the Italian – Dutch Beppo SAX satellite. The XRT light curves show common characteristics that are reasonably understood within the framework of the fireball model. Unforeseen flares are however detected in a large fraction of the GRB observed and the energy emitted by the brightest ones may be as much as 85% of the total soft X ray emission measured by XRT. These characteristics seems to be common to long and short bursts.

## 1. INTRODUCTION

As discussed by Gehrels in these proceedings the Swift satellite (Gehrels et al. 2004), is giving us the answer to many of the questions that remained unsolved following the Beppo-SAX mission. The soft X_ray light curves (0.2 – 10 keV) have been observed with great detail and temporal resolution for very long periods, the record being detained by GRB050408 that was observed over a period of 38 days (Capalbi et al. 2005). More important the rapid pointing capabilities of the spacecraft allows the re-pointing of the target in some cases in less than 60 seconds after the alert to immediately start the observation in the X ray band and in the optical (1700 Å to 6500 Å). The accurate determination of the X ray position with the X ray Telescope (XRT) allowed also the localization of the short Gamma Ray Bursts. We have discovered that these occur on the outskirt of rather evolved galaxies, early types or $E_m$+A galaxies, and are accompanied by an isotropic emission that is about a factor 100 – 1000 smaller that the isotropic energy emitted by the long bursts. The shorts were all detected in the proximity of nearby galaxies with $z < 0.5$. The low redshift of the sample may in part be due to the lower energy emitted and to the difficulties of the follow up observations but mainly related to the star formation history of the Universe (Guetta & Piran 2005). In some cases the afterglow could also be absent if the event is located at the very outskirt of the galaxy where the ISM has an extremely low density. These observations are also opening our research toward a better understanding of the stellar evolution of massive stars and merging of relativistic stars.

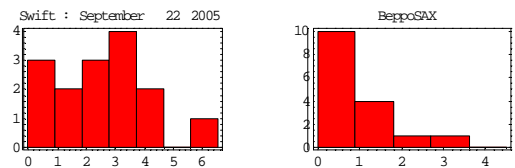

Fig. 1. The distribution of the redshifts of the GRB discovered by the Swift and by the Beppo SAX satellites. While the statistics is still small we have a strong indication that Swift tends to detect a larger number of high z objects.

Perhaps the most unexpected feature, this is not yet fully understood, is the capability of Swift to detect objects at high redshifts. As illustrated in Fig. 1 the mean redshift of the sample so far observed by Swift is quite larger than the mean redshift observed by other missions. This may be due to the fact that in the past the faint tail of the distribution or the rapid decaying afterglows were missed and to the band pass and high sensitivity of BAT. But most important the hope to detect objects at very high z was satisfied with the detection of GRB050904 for which z = 6.29

(Cummings et al., 2005, Cusumano et al. 2005, Tagliaferri et al. 2005). This detection is of paramount importance since it shows that the fast evolution of a massive star producing a GRB occurs also at epochs where stars and galaxies have very low metal abundance. More interesting we now know that we have the capability to detect these objects near the re-ionization epoch soon after the Universe exits from the Dark Age and likely we may get the possibility to learn about the formation and evolution of the early stars assuming a gamma ray burst can still occur when the composition is more or less primordial. This detection gave new impetus toward more ambitious goals and toward the developments of new strategies and computations. We believe, therefore, that the discoveries made by Swift opened up a very broad spectrum of research supplying a unique data base and new perspectives.

Previous work carried out with Beppo SAX and other satellites already evidenced that the basic model as proposed by Meszaros and Rees (1993), see however also alternative models which have been proposed, was satisfactory. For details see the excellent reviews by Piran (1999), Hurley et al. (2002), Zhang and Meszaros (2004). The observations by Swift confirmed the validity of the models and however the main challenge now is the estimate of whether or not there are constant patterns in the light curve indicating that not only we are dealing with the same kind of physics but that the evolution and magnitude of the phenomenon follows certain well understood patterns. In this context the shape of the light curves in the rest frame, their characteristics and the energy balance, especially when compared to the energy related to the prompt emission, are highly significant in refining the models. In section II we are dealing with the first set of rest frame GRB light curves, and related energy, that we observed before May 15 this year. In section III we deal with the morphology of flares and the energy involved. In particular we show that long burst flares, short bursts flares, low z flares and high z flares have all some common characteristics in the light curve. Flares that are visible during the late phases of the light curve give evidence of long lasting activity of the central engine. Turning around the problem it remains to explain, among other things, why the flares are not present in all the bursts. Unless otherwise stated we use $H_0 = 65$ km s$^{-1}$ Mpc$^{-1}$, $\Omega_m = 0.3$ and $\Omega_\Lambda = 0.7$.

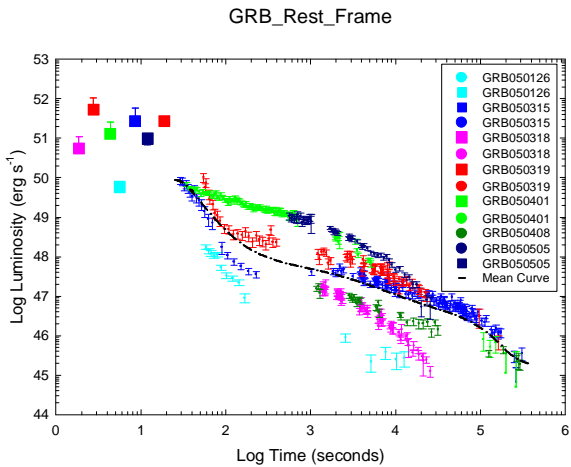

Fig. 2. The light curves have been plotted in the rest frame of each single burst. For the origin of time we used the trigger time as given by the BAT instrument. Squares refer to the mean flux observed by BAT during the burst and converted via the BAT spectrum to the energy band of XRT, circles refers to the observed flux in the band 0.2 – 10 keV. The dot dashed line is the mean curve as described in the text.

## 2. REST FRAME LIGHT CURVES.

By the 15$^{th}$ of May 2005 we had a set of only 7 XRT light curves for which also the redshift was available and none of these had flares (see section III). Flares had been already detected however in GRB050406 and in GRB050502B (Burrows et al. 2005, Romano et al. 2005 and Falcone et al. 2005). From these no flare light curves (Chincarini et al. 2005), it was immediately clear that we had at least two types of basic morphology. The first type consisting of light curves presenting a very sharp decay (power law exponent generally between -3 and -2), at the very beginning of the observations (about 30 seconds after the BAT trigger in their rest frame) followed by a mild slope (power law exponent around -0.6 with rather large variations) that would later steep (power law exponent < -1). The second type, vice versa, would start with a mild slope and later steep in a way similar to the last variation observed in the previous type (note however that GRB050401 likely shows one mini-flares). The flares observed in many light curves in the following months are simply superimposed to this basic morphology. The findings of the early sample have been confirmed, see also Nousek et al. (2005). The expected relation between the temporal decay index and spectral index is universal for synchrotron emission from spherical fireballs (or jets with an opening angle much larger than the relativistic beaming scale), and is given by $\alpha = 2 + \beta$ (Kumar and Panaitescu 2000) where $F_\nu \propto t^{-\alpha} \nu^{-\beta}$. The typical value of $\beta$ is about 0.5 -1., the maximal decay index could be $\alpha \sim 3$. If the decay is steeper than 3, as it may be the case for GRB 050319, we possibly have to argue for a highly collimated jet, and however in this case the time adopted for the trigger may be wrong (Chincarini et al. 2005, Barthelemy et al. 2005).

The prompt emission by GRB050319 is characterized by two bursts separated by about 137 seconds (see Fig. 2) while in GRB050401 we observe two burst separated however only by a few seconds. Likely the true trigger [$T_0$] for the afterglow of GRB050319 is related to the second burst since the effects due to the first burst had enough time to decay and be overshadowed by the afterglow of the second burst.

This may not be the case for GRB050401 since the two bursts detected by BAT are one immediately after the other. In this case however note that the formal BAT zero time is more or less half way of the first burst, and this matter must be clarified as well. An other possibility exists that the steep decay of the first phase (30 to 100 seconds in the rest frame) is due to the decay curve of a flare whose peak was missed. This possibility can not yet be completely excluded. However it seems unlikely due to the nearby similat slope shown by most GRBs. Flares, on the contrary, show decay slopes that differ largely from flare to flare. Likewise it is unlikely that the type II is the consequence of a missed flare or of a missed phase I type I curve. In the case of GRB050401, but now we have a much larger sample, the XRT observations started about 30 seconds (rest frame) after the BAT trigger and we should have seen the sign either of a late prompt emission or of a flare. We conclude, base on the evidence we have today, that the X-ray light curves of GRB are divided mainly in two types. In addition many GRB show flare superimposed to the light curve.

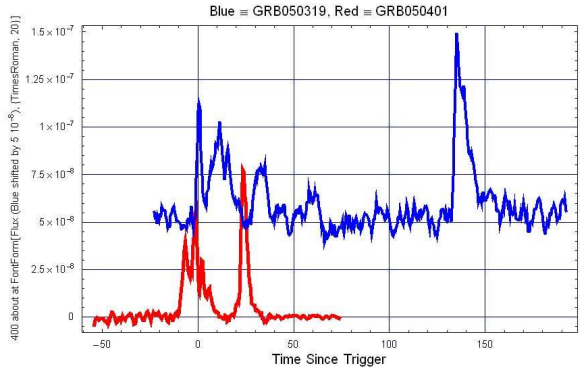

Fig. 3. The BAT light curve of GRB050319 and GRB050401 with the background of GRB050401 shifted up by 5 10$^{-8}$ (erg cm$^{-2}$ s$^{-1}$) for clarity.

For each light curve we measured both the mean spectrum and the spectrum before and after the light curve breaks to look for spectral variations. The energy index of the spectrum does not change much (the error of the mean is larger however that the error of a single estimate) from burst to burst and does not change crossing the break. We measure a mean value of the Energy Index (EI) = 1.12 ± 0.31. The rather large error is due mainly to GRB050319 that shows variation in the EI as a function of time. The hardness ratio in general changes during the flares in a way that is similar to the flare light curve but with a phase lag (this work is in preparation).

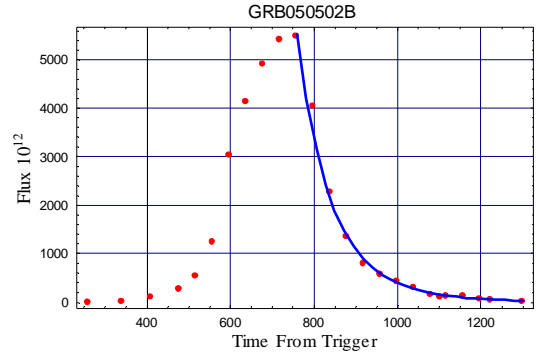

Fig. 4. Profile of the GRB050502B giant flare. The flux is expressed in 10$^{-12}$erg cm$^{-2}$ s$^{-1}$ units and after subtraction of the "background" light curve. The slope of the decay changes from -9.58 to -7.12 if rather than using for the origin of time the trigger by BAT we use the beginning of the flare itself.

The ratio of the energy emitted during the afterglow (or the Total Fluence) to the energy emitted during the prompt emission changes considerably from Burst to Burst and is in the range, for the sample of Fig. 2, from 0.016 to 0.40. For the long bursts we observe in the soft X ray afterglow at most 40% of the energy emitted during the prompt emission. An other interesting and more accurate way to compute the relation between prompt emission and afterglow is to use well defined time interval of the afterglow and compute the rest frame Fluence. Using the rest frame intervals 50 – 200 seconds and 1300 – 12600 seconds we find a tight correlation between the prompt emission and the rest frame afterglow emission indicating a rather tight correlation between the energy injected in the ISM, within the external shock scenario, and the energy emitted in the soft X ray during the afterglow.

## 3. FLARES: MORPHOLOGY AND ENERGETIC

The morphology of the flares, likely the direct consequence of secondary bursts, have a rather simple morphology at least in those cases where the flares are not clustered and the signal to noise is high. Subtracting the "background" decaying light curve, that in the cases we considered has the morphology described in section II, the shape is characteristic of the emission due to the clash between to relativistic shells as described in Kobayashi et al. (1997), that is a rapid rising power law followed by a very rapid power law decay. In the computation of the flare decay slope rather than using the BAT trigger time we should use an origin of time coincident with the beginning of the flare. In the case of GRB050502B the exponent of the power law decay changes, by shifting the origin of time by about 350 seconds, from -9.58 to -7.12. The change is highly significant even if it does not cause any change that could drastically modify the possible models. The energy emitted during the flare is comparable, in some cases, to the energy emitted during the whole XRT afterglow and a rather large percentage of the prompt emission energy. The prompt emission fluence as measured by BAT is $F_{BAT}=$ 4.7 $10^{-7}$ erg cm$^{-2}$ while $F_{XRT}=$ 1.7 $10^{-6}$ erg cm$^{-2}$, $F_{Flare}$ (750 s) = 1.43 $10^{-6}$ erg cm$^{-2}$ and $F_{Flare}$ (75000 s) = 2.21 $10^{-7}$ erg cm$^{-2}$ The first big flare has a Fluence that is about 85% of the whole XRT light curve Fluence and larger than that of the prompt emission. The flares can not be a consequence of the initial trigger but rather due to new bursts in the central engine.

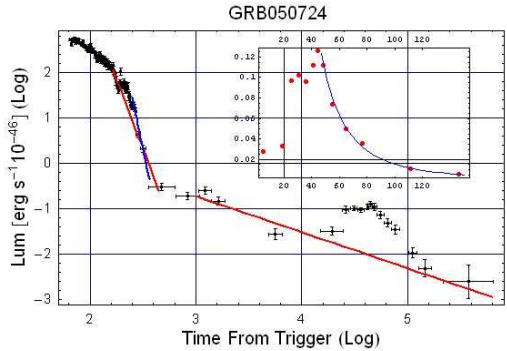

Fig. 5. The morphology of the late flare in the short GRB050724 is exactly the same as the morphology of the flare observed in long bursts and characteristics of the clash of relativistic shells. The slope of the decaying part of the flare, after subtracting the main light curve – red line in the figure, is $\alpha = -2.75$ where the time is referred to the BAT trigger time. Note that in the inset graph, linear coordinates, the time is given in seconds/1000 and the Luminosity (ordinates) in $10^{-46}$ erg s$^{-1}$. The following data refer to the rest frame: BAT (15 -350 keV, $\Delta t = 0.79$ s) emission = 1.1 $10^{50}$ erg; XRT (0.2-10 keV, $\Delta t = 65 – 376456$ s)=4.1 $10^{50}$ erg, first flare at about 220 s RF_Fluence = 1.47 $10^{49}$ erg, large flare at about 50000 s RF_Fluence = 6.03 $10^{49}$. For more details see Campana et al. in preparation.

What is striking is that the same flare structure has been observed in the long burst at high z, GRB050904 (Tagliaferri et al. 2005, Cusumano et al. 2005 and references therein) and in the only short burst, GRB050724, for which we were able to observe a detailed X ray light curve. In Fig. 5 we show in the inset the late flare with its power law decay. These observations show that flares may occur in any type of burst at any time along the light curve. This flare at about 40000 seconds (about ½ a day) clearly show that in the short bursts (as we have seen in the long bursts) the central engine must remain active for a long time and that a similar mechanism must be at work. The decay time is rather long.

This is striking if we consider the differences between the progenitor and the parent population of short and long bursts. While long bursts are the end result of the fast collapse of a very massive star (Woosley S.E. 1993), and these are generally located in star forming late type galaxies, Bloom et al. (2002), Le Floch et al. (2003) evidence is now building up that the progenitors of the short bursts are merging relativistic binaries clearly generated by an older parent population. The break through came with the Swift observations of GRB050509B (Gehrels et al. 2005), since for the first time it has been possible to get an accurate position, based on a total of 11 photons we got in the first 1640 seconds of integration time, with the XRT telescope which started to acquire data 62 seconds after the trigger. The observations we obtained with the VLT (but we had also unpublished images with the TNG), see Fig. 1 in Gehrels et al. (2005), had in the XRT error circle the outskirt of a E1 galaxy (likely the host of the short GRB) and a large number of faint background galaxies. The E1 being the II brightest galaxy of a cluster of galaxies. This interpretation seemed to fit with the coalescence model of relativistic stars (NS-NS, NS – BH). GRB050724 fully confirmed this interpretation with the detection of the afterglow near a rather bright elliptical galaxy (Barthelemy et al. 2005, Berger et al. 2005).

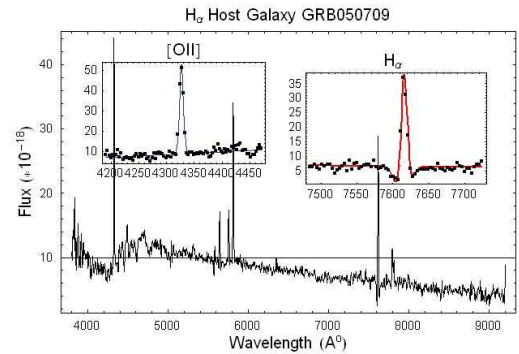

Fig. 6. Spectrum of the Host Galaxy of GRB050709. The star formation rate derived from $H_\alpha$ emission: SFR=0.13 $M_\odot$ / yr. The Balmer line clearly show the emission in the center of a rather broad absorption line. The [OII] nebular doublet is shown here only for comparison.

GRB050709 has been detected by HETE and the optical afterglow imaged by Hijorth et al. (2005) and later by Covino et al. (2005) with the ESO VLT. In this case the host galaxy is a blue dwarf irregular (Fox et al. 2005), and the spectrum (Covino et al. 2005), shows evidence of star formation activity, emission lines are clearly detected. Most interesting there is also evidence of a stellar population of type A as clearly indicated by the Balmer absorption line visible in the spectrum, Fig. 6. This means that the galaxy had a starburst about 5 $10^9$ ago. It is misleading to associate the presence of the short GRB to a star forming galaxy but rather it is associated with an old starburst the galaxy had long ago. In other words we do not expect in the galaxy population of the GRB050709 the presence of very massive stars. We conclude that the long and short bursts have different progenitors and different parent population and occur in different environments.

The high z GRB050904 has been observed with XRT immediately after the trigger and, as shown in Fig. 7,

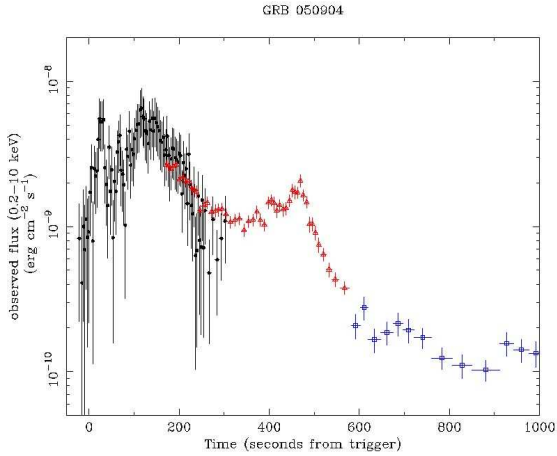

Fig. 7. The first set of observations have been obtained with the BAT telescope while the flare, second burst, has been detected by the XRT Telescope. The slope of the power law decay in the rest frame is α = -2.14.

we have an important overlap between the data obtained by BAT and those obtained by XRT. It may clarify, but we will have more statistics in the following months, the morphology of the Type I light curves. The BAT light curve continue smoothly into the XRT light curve and the rest frame slope is in agreement with the slope generally observed in the soft X ray afterglow, $\propto t^{-2.14}$ in the rest frame. There is continuity between BAT and XRT (BAT is sensitive down to the 15 – 25 keV band, what we generally call the soft channel band) and simply shows how the internal shock decays and continue with the emission of the afterglow (external shock). On the other hand it is not clear when exactly is the external shock emission beginning.

For completeness we also show in Fig. 8 the GRB050904 flare. Whether or not the flare activity is different in high z flares, at z > 6 we expect to have stars and ISM with a very low metal abundance, as to be determined. For this only GRB we observed at very high z a peculiarity my be the very high flare activity we observe during the mild decay following the first phase of the light curve where the slope is sensibly higher.

To summarize the morphology we discussed for long bursts, morphology that seems to be similar to that of the light curves of short bursts, referring to GRB050724 that however is the only case observed and has a type I morphology and missed break, we reproduce the light curve of GRB050822 that include all of the characteristics we have described.

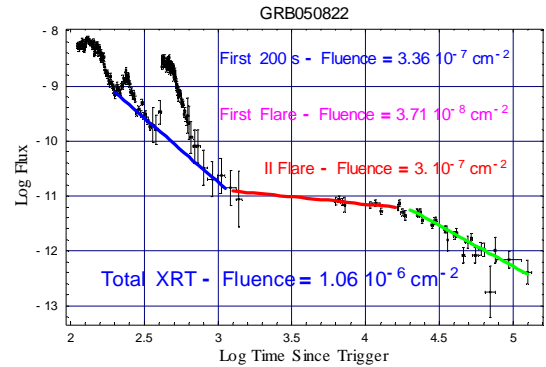

Fig. 9. Following the early maximum the light curve shows a first phase, slope -2.34, followed by a milder decay, slope -0.27 steeping to a -1.48 slope in the late phase. Superimposed to this type I light curve we have two big flares.

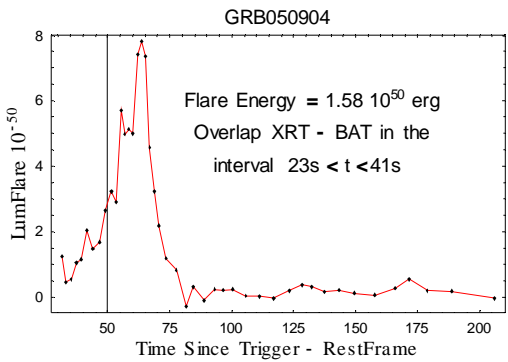

Fig. 8. The First Flare, Luminosity versus time, of the XRT light curve of GRB050904, see Fig. 6, in rest frame coordinates. Note also in this case the rapid decay and however a less rapid rise to maximum that could be due irregular energy injection however.

## ACKNOWLEDGMENTS.

We would like to thank Dino Fugazza for helping with the manuscript. The work is supported in Italy by funding from ASI on contract number I/R/039/04, at Penn State by NASA contract NAS5-00136 and at the University of Leicester by PPARC contract number PPA/G/S/00524 and PPA/Z/S/2003/00507. We acknowledge in particular all those member of the Swift Team at large who made this mission possible. This goes from the building of the hardware, the writing of the software, the operation at the Mission Operation Centre and the performance of the ASI ground segment at Malindi, Kenya.